# Evidence of a two-dimensional nitrogen crystalline structure on silver surfaces

Xuegao Hu,[1,2,#] Haijun Cao,[1,2,#] Zhicheng Gao,[1,2,#] Hui Zhou,[1,2,#] Daiyu Geng,[3] Dong Li,[1,2] Jisong Gao,[1,2] Qiaoxiao Zhao,[1,2] Zhihao Cai,[1,2] Peng Cheng,[1,2] Lan Chen,[1,2*] Sheng Meng,[1,2*] Kehui Wu,[4,5*] Baojie Feng,[1,2,5,6*]

[1]*Institute of Physics, Chinese Academy of Sciences, Beijing, 100190, China*
[2]*School of Physical Sciences, University of Chinese Academy of Sciences, Beijing, 100049, China*
[3]*Max Planck Institute of Microstructure Physics, Weinberg 2, 06120 Halle, Germany*
[4]*Tsientang Institute for Advanced Study, Zhejiang 310024, China*
[5]*Interdisciplinary Institute of Light-Element Quantum Materials and Research Center for Light-Element Advanced Materials, Peking University, Beijing, 100871, China*
[6]*Songshan Lake Materials Laboratory, Dongguan, Guangdong, 523808, China*

[#]These authors contributed to this work.
[*]Corresponding author. E-mail: lchen@iphy.ac.cn; smeng@iphy.ac.cn; khwu@tias.ac.cn; bjfeng@iphy.ac.cn.

**Nitrogen, the most abundant element in Earth's atmosphere, exists as a diatomic gas under standard temperature and pressure. In the two-dimensional (2D) limit, atomically thin nitrogen, termed nitrogene, has been theoretically predicted to form crystalline materials with various polymorphic configurations, exhibiting diverse chemical and physical properties. However, the synthesis of nitrogene has remained elusive due to the strong nitrogen-nitrogen triple bonds. Here, we report experimental evidence of the formation of nitrogen-based crystalline structures compatible with nitrogene on silver surfaces via ion-beam-assisted epitaxy. Through a combination of scanning tunneling microscopy, angle-resolved photoemission spectroscopy, and first-principles calculations, we demonstrate that**

**the nitrogene-like structure adopts a puckered honeycomb lattice. Notably, our calculations predict a nitrogene band gap of up to 7.5 eV, positioning it as a promising candidate for ultraviolet optoelectronic devices and high-k dielectric applications.**

Introduction

Synthetic two-dimensional (2D) materials without layered parent allotropes exhibit a remarkable diversity of physical and chemical properties due to their structural variability. A prominent class of such materials is the Xenes, which include borophene[1,2], silicene[3,4], stanene[5,6], tellurene[7,8], molybdenene[9], and goldene[10]. Most Xenes identified to date are semimetals or metals, which limits their potential applications in logic electronics and optoelectronics. To address this limitation, pnictogen-based Xenes, such as phosphorene[11-13], antimonene[14,15], and bismuthene[16-18], have garnered significant attention due to their semiconducting properties and promising device applications[19,20], as well as their range of exotic properties. For instance, black phosphorene field-effect transistors exhibit excellent drain current modulation and high carrier mobility[12], while honeycomb bismuthene is a quantum spin Hall insulator with a substantial inverted band gap[16]. Additionally, black phosphorus-like bismuthene has been identified as an elemental ferroelectric material[18].

As the lightest pnictogen, nitrogen predominantly exists as diatomic molecules in nature, which are gaseous under standard temperature and pressure. The strong N≡N triple bonds impart high stability and chemical inertness to molecular nitrogen. In contrast, crystalline nitrogen can usually be realized under extreme conditions of low temperature and high pressure, including cubic-gauche polymeric nitrogen (cg-N)[21], black phosphorus-like nitrogen (bp-N)[22,23], and hexagonal layered polymeric nitrogen (HLP-N)[24]. However, the pure phases of these crystalline structures decompose upon pressure release and cannot exist under ambient conditions, although cg-PN can be stabilized in a composite environment under near-ambient conditions[25].

The 2D crystalline monolayer of nitrogen, termed nitrogene, has been theoretically proposed to adopt various polymorphic structures, including the honeycomb[26-28], black phosphorus-like[29,30], square-octagon[31-33], and zigzag sheet structures[34,35]. Its strong covalent bonding and high chemical reactivity make nitrogene a promising candidate for applications in field-effect transistors, catalysis, gas storage, and solar cells. However, the synthesis of nitrogene remains challenging due to the significant energy required to break the strong N≡N triple bonds and the need to select an appropriate substrate for epitaxial growth.

Here, we report the synthesis of a nitrogene-like structure on a Ag(100) surface via ion-beam-assisted epitaxy (IBAE). Scanning tunneling microscopy (STM) and low-energy electron diffraction (LEED) measurements reveal that the nitrogene-like structure forms a ($\sqrt{2}\times\sqrt{2}$)R45° superstructure relative to the Ag(100) substrate. In contrast to other Xenes, it exhibits a wide electronic band gap (~7.5 eV), as confirmed by angle-resolved photoemission spectroscopy (ARPES), scanning tunneling spectroscopy (STS), and density functional theory (DFT) calculations. Based on these results, we propose a puckered honeycomb structure for the 2D nitrogen sheets, which is consistent with all experimental observations. The successful synthesis of the nitrogene-like structure not only offers insights into the bonding behavior of crystalline nitrogen but also paves the way for the development of nitrogene-based optoelectronic and plasmonic devices.

**Results**

**Synthesis of nitrogene-like structures on Ag(100)**

The growth mechanism of nitrogen on Ag(100) via IBAE is illustrated in Fig. 1a. Molecular $N_2$ is chemically inert due to its strong N≡N triple bond (~10 eV), which prevents it from adsorbing on substrates. To address this challenge, we employed an ion source operating with pure $N_2$ gas to generate reactive nitrogen species, primarily $N_2^+$ ions, which also provide sufficient kinetic energy for the ions to reach the Ag(100) surface. During growth, the kinetic energy of the $N_2^+$ ions is crucial for breaking the

N≡N triple bonds through collisions with Ag atoms. Meanwhile, the remaining energy of the cracked N atoms can increase atomic mobility on the surface and improve the quality of the sample. However, excessive kinetic energy can damage the Ag(100) surface, leading to surface corrugation. Through optimization experiments, we determined that the ideal kinetic energy required for the synthesis of the nitrogene-like structure on Ag(100) is approximately 30 eV.

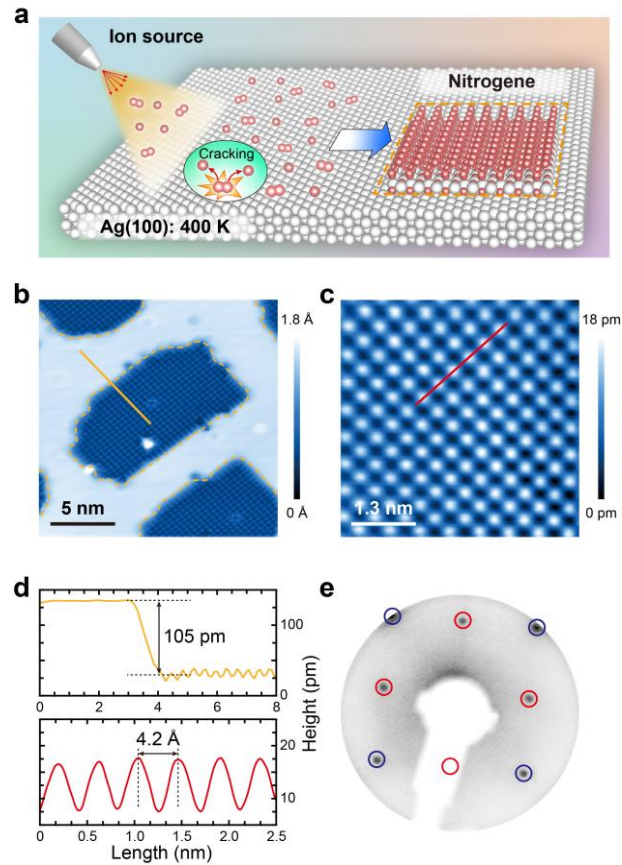

**Fig. 1| Growth of nitrogene-like structures on Ag(100). a**, Schematic illustration of the growth process. Low-energy nitrogen ions are generated by an ion source and bombarded onto the Ag(100) surface. **b**, Large-scale Scanning tunneling microscopy (STM) image (V=-1 V, I=100 pA). Nitrogen-covered areas are outlined with orange dashed lines. **c**, High-resolution STM image of the nitrogene-like structure on Ag(100) (V = 0.06 V, I = 100 pA). **d**, Line profiles along the orange dashed line in panel **b** (top) and the red line in panel **c** (bottom). **e**, low-energy electron diffraction (LEED) pattern of the nitrogene-like structure on Ag(100), showing a Ag(100)-($\sqrt{2}\times\sqrt{2}$)R45° superstructure. Blue and red circles indicate diffraction spots corresponding to Ag(100) and the nitrogene-like structure, respectively.

The substrate temperature was maintained within a narrow range of 400±10 K during the growth process. At lower temperatures, disordered nitrogen clusters formed on the surface, while higher temperatures led to nitrogen desorption. For comparison, we performed growth under the same $N_2$ pressure but with the ion source turned off. In this case, no nitrogen remained on the surface, indicating that the generation of $N_2^+$ ions is essential for the successful growth of the nitrogene-like structure.

Figure 1b presents a large-scale STM image of the nitrogene-like structure on the Ag(100) surface (see also Supplementary Fig. 1). The apparent height of the islands is slightly lower than that of the adjacent Ag(100) surface, likely due to the removal of the topmost Ag layer during growth by the chemically assisted reorganization process. This phenomenon is commonly observed in systems involving foreign atoms on metal surfaces, such as the Si/Ag interface[36,37]. Another reason for the recessed feature is the insulating nature of nitrogene, which results in a lower apparent height in constant-current STM imaging. The islands exhibit a well-ordered atomic structure, as shown in the high-resolution STM image in Fig. 1c and the line profile in Fig. 1d. These images reveal that the nitrogene-like structure adopts a square lattice with a lattice constant of 4.2 Å. The primitive lattice vectors of the nitrogene-like structure are rotated by 45° relative to those of the Ag(100) substrate, resulting in a ($\sqrt{2}\times\sqrt{2}$)R45° superstructure. This superstructure is further confirmed by the LEED measurements shown in Fig. 1e. Moreover, systematic LEED measurements confirm the formation of a single structural phase on Ag(100), as shown in Supplementary Fig. 2.

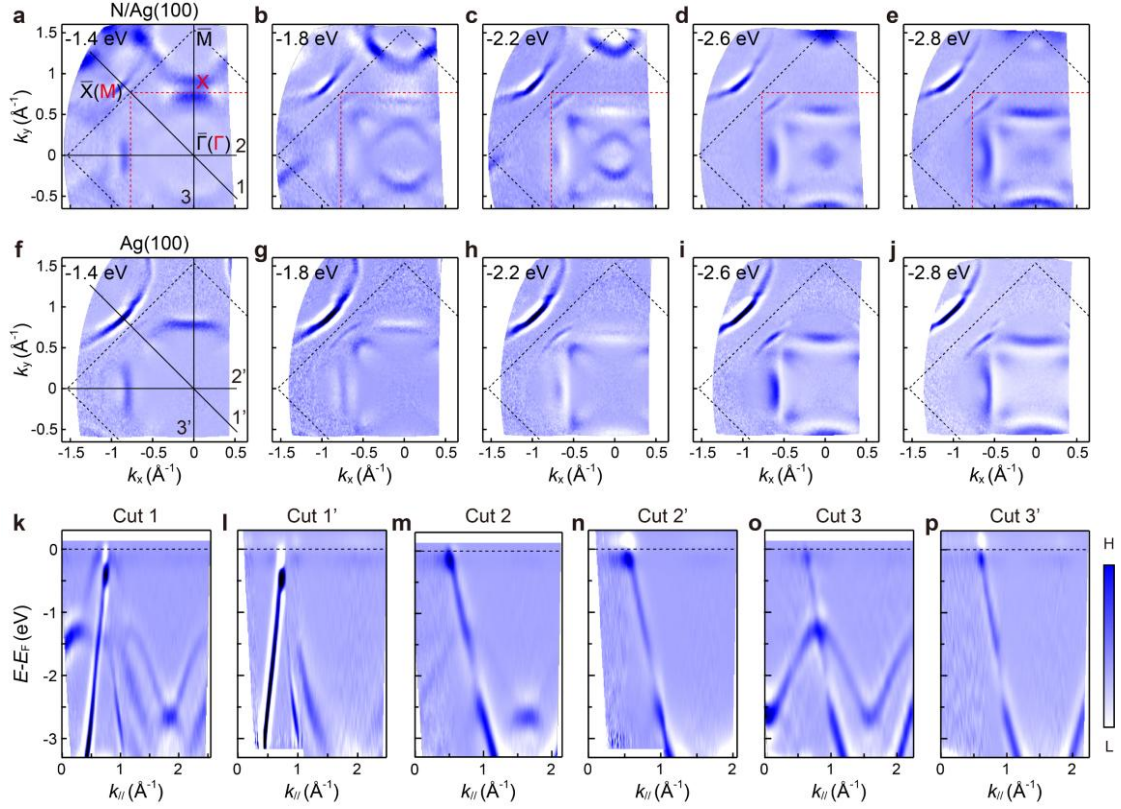

**Fig. 2| Angle-resolved photoemission spectroscopy (ARPES) measurements of the nitrogene-like structure on Ag(100). a-e**, Constant energy contours (CECs) at binding energies of 1.4, 1.8, 2.2, 2.6, and 2.8 eV, respectively. The black and red dashed lines indicate the first Brillouin zones (BZs) of Ag(100) and the nitrogene-like structure, respectively. The $\bar{X}$ point of Ag(100) coincides with the M point of the nitrogene-like structure. **f-j**, CECs of pristine Ag(100) at the same binding energies as panels (a–e). **k-p**, ARPES intensity plots along Cuts 1-3 and 1'-3', as marked in panels **a** and **f**. Cuts 1-3 correspond to the nitrogene-like structure on Ag(100), while Cuts 1'-3' correspond to pristine Ag(100). All CECs and cuts are presented as second-derivative images.

## ARPES measurements

To investigate the electronic structure of the nitrogene-like structure, we performed ARPES measurements. Figures 2a-2e display the constant energy contours (CECs) at various binding energies, while Figures 2f-2j show the corresponding CECs for pristine Ag(100) for comparison. No electronic states associated with the nitrogene-like structure were observed within 1.3 eV below the Fermi level, indicating a semiconducting or insulating nature of the overlayer. At a binding energy of approximately 1.4 eV, a prominent circular feature appears at the Γ point. As the binding energy increases, this feature gradually shrinks into a dot-like shape at 2.6 eV, then expands back into a circular feature at 2.8 eV, indicating the presence of both electron-like and hole-like bands. Notably, similar features are also observed at higher-order Γ points, consistent with the periodicity of the nitrogene-like structure.

ARPES intensity plots along Cuts 1-3 and 1'-3', as indicated in Figs. 2a and 2f, are shown in Figs. 2k-2p. Along the Γ-M direction of the nitrogene-like structure (Cut 1, Fig. 2k), both an electron-like band and a hole-like band are observed at the Γ point, nearly touching each other at ~2.5 eV below the Fermi level. An parabolic fit of the electron-like bands yields an effective mass of approximately $0.5m_0$ (see Supplementary Fig. 3) Cuts 2 and 3 are both along the Γ-X direction, but with parallel and perpendicular orientations relative to the slit of the electron energy analyzer, respectively. In Cut 2 (Fig. 2m), the lower hole-like band is more pronounced, while in Cut 3 (Fig. 2o), the upper electron-like band is more intense. These variations in spectral intensity are likely due to photoemission matrix element effects. Importantly, no similar bands are observed in pristine Ag(100), regardless of momentum or energy shifts, as shown in Figs. 2l, 2n, and 2p. This excludes the possibility of band-folding effects or rigid shifts of the bulk bands.

**First-principles calculations**

We performed first-principles calculations to explore potential structural models of the nitrogene-like structure on Ag(100). The most likely structural model is shown in Figs. 3a and 3b. In this model, the N atoms in the nitrogene layer exhibit buckling at different heights, forming a black-phosphorus-like structure. The slight difference from the black phosphorus structure might be attributed to the substrate-induced relaxation. An alloyed buffer layer, composed of AgN in a stoichiometric ratio, exists between the nitrogene layer and the Ag(100) substrate. This buffer layer serves to passivate the metal surface, thereby facilitating the formation of the nitrogene overlayer. Such phenomena are commonly observed in the growth of other 2D pnictogens[38-40]. Notably, the buffer layer is unstable on its own and requires the presence of the nitrogene layer for stabilization, as demonstrated in Supplementary Fig. 4.

Figure 3c presents the top view of the structural model, along with experimental and simulated STM images, in agreement with each other. The bright protrusions in the STM images correspond to the topmost N atoms in the model, as indicated by the red balls. The existence of a buffer layer is consistent with our XPS measurement results, as shown in Fig. 3d and 3e. We observe two N 1*s* peaks, which correspond to the N atoms in the buffer layer and the nitrogene layer, respectively.

We further calculated the formation energies per atom of freestanding nitrogene and $N_2$ molecules, obtaining values of −8.047 eV and −8.304 eV, respectively. Although nitrogene is thermodynamically less stable than molecular nitrogen in the freestanding form, it can be stabilized on the Ag(100) substrate through interaction with the buffer layer, enabling its experimental realization under our growth conditions.

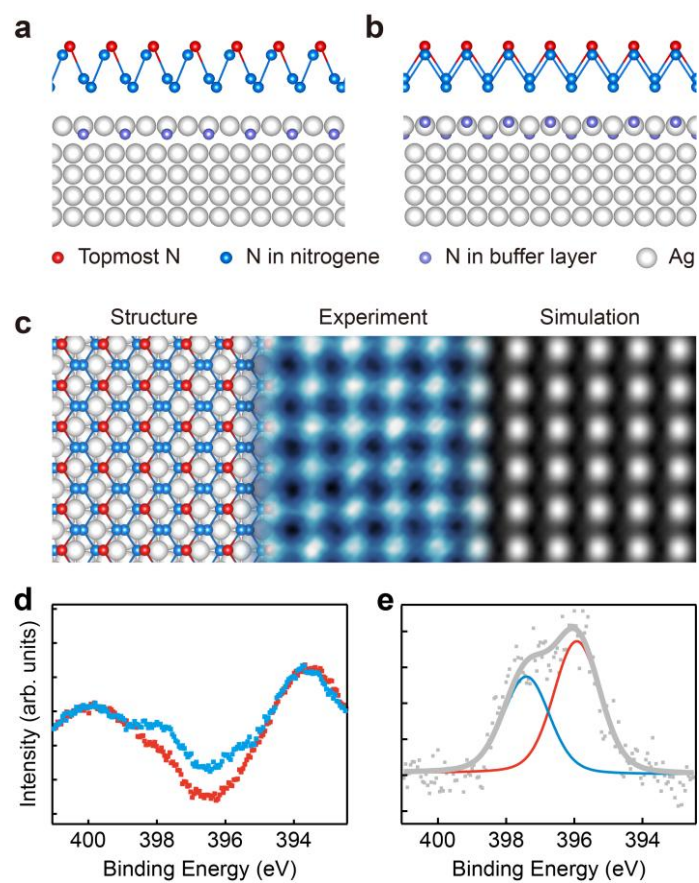

**Fig. 3| Atomic structure of the nitrogene-like structure on the Ag(100) surface. a,b**, Side view of nitrogene on Ag(100). Nitrogen atoms are colored according to their height for clarity. A buffer layer with a stoichiometric composition of AgN exists between the nitrogene layer and the Ag(100)

substrate. **c**, (Left to right) Top view of the structural model, experimental STM image (V=1 V, I=100 pA), and simulated STM image by the density functional theory (DFT). Bright protrusions in the STM images correspond to the topmost nitrogen atoms in the structural model, highlighted in red. (d) X-ray photoelectron spectroscopy (XPS) spectra of the nitrogene-like structure grown on Ag(100) (blue) and pristine Ag(100) (red). (e) Difference spectrum obtained by subtracting the pristine Ag(100) signal (red) from the nitrogen/Ag(100) spectrum (blue) shown in panel (e).

Figures 4a and 4b display the ARPES spectra and calculated band structures along the M-Γ-M direction, respectively. There is good agreement between the experimental and calculated band structures, as highlighted by the orange rectangles and red dashed lines in Figs. 4a and 4b. Figure 4c shows the calculated density of states (DOS) for the system. A peak is observed at 2-3 eV below the Fermi level, corresponding to the touching points of the electron and hole pockets in the ARPES spectra. Above the Fermi level, a prominent peak at approximately 1.1 eV is visible in the DOS, which aligns with our dI/dV spectra shown in Fig. 4d. These results unambiguously confirm the proposed structural model for the nitrogene-like structure on Ag(100).

The calculated band structures of freestanding nitrogene are shown in Supplementary Fig. 5. Notably, the conduction bands of freestanding nitrogene are identical to those calculated for nitrogene on the substrate, indicating weak hybridization between nitrogene and the buffer layer. It should be noted that the interaction between nitrogene and the buffer layer is still strong enough to prevent lateral displacement or tip-induced manipulation of the nitrogene islands. The calculated band gap of freestanding nitrogene is approximately 7.5 eV, the largest among experimentally realized 2D wide-gap materials, such as h-BN. The valence bands are located around 5 eV below the Fermi level, which are masked by the strong spectral weight of the bulk bands from Ag(100), as shown in Supplementary Fig. 6.

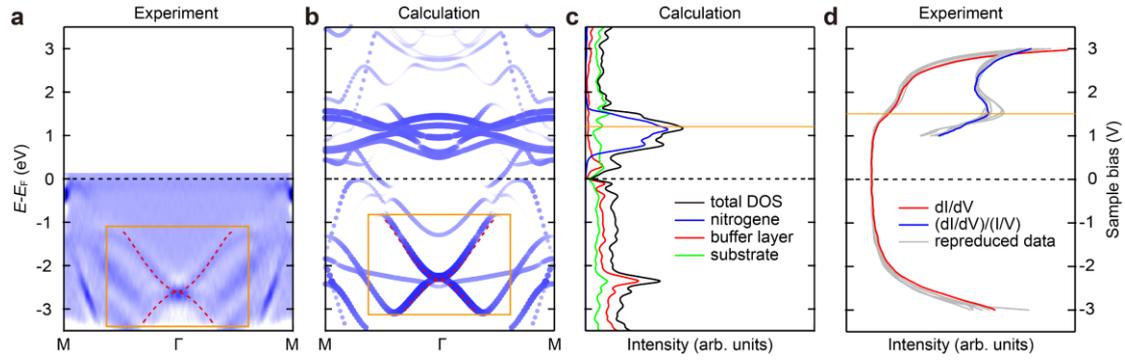

**Fig. 4| Electronic structures of the nitrogene-like structure on Ag(100). a**, ARPES intensity plot along the M-Γ-M direction. The shapes of the energy bands are highlighted with red dashed lines as guides to the eye. **b**, Calculated band structures along the same momentum path as in **a**. The calculated bands agree well with the ARPES results, with key features highlighted by orange rectangles in both panels **a** and **b**. **c**, Calculated density of states (DOS) within the same energy range, showing a prominent peak at 1.1 eV above the Fermi level, marked by the orange line. **d**, dI/dV spectra measured within the corresponding bias range. The normalized differential conductance, *i.e.*, dI/dV/(I/V), are overlaid to enhance the visibility of the peak feature.

## Discussion

Finally, we discuss alternative structural models for the N/Ag(100) system considered during our structural search. Due to constraints from lattice constants and symmetry, only a limited number of structural models are feasible, as shown in Supplementary Fig. 7 to Fig. 9. Several criteria were employed to identify the correct model: first, it must remain stable during structural optimization; second, the simulated STM images, calculated band structures, and DOS must all align with the experimental observations. After an extensive and systematic search, only the structure shown in Fig. 3 satisfies all these criteria. However, unambiguous confirmation of the precise structural model of the N/Ag(100) system will require further experimental and theoretical investigation.

The synthesis and characterization of the nitrogene-like structure reveal not only its unique structural and electronic properties but also provide insights into the intriguing chemical behavior of nitrogen in the 2D limit. While pure nitrogen typically exists as diatomic $N_2$ molecules, stabilized by the strong N≡N triple bond and chemically inert under standard conditions, the nitrogene lattice stabilizes nitrogen atoms in a crystalline structure with dispersing electronic bands. This represents a significant deviation from the molecular nitrogen phase. The structural transformation suggests that 2D nitrogen can exhibit properties distinct from its gaseous counterpart, including enhanced catalytic reactivity and tunable bonding environments. These findings underscore the chemical versatility of nitrogen and open promising avenues for its application in energy storage, catalysis, and electronic devices. Notably, the nitrogene-like structure is stable at room temperature during both synthesis and characterization, which facilitates device integration and broadens the range of practical environments where nitrogene-based materials may operate. In addition, the large band gap of nitrogene (~7.5 eV) positions it as a potential candidate for applications in ultraviolet optoelectronics, where materials with wide band gaps are crucial for efficient light emission, detection, and sensing. On the other hand, the large band gap, combined with its 2D nature, makes nitrogene a promising candidate for high-*k* dielectric materials in semiconductor devices, potentially enabling faster, more energy-efficient electronics. The combination of the exotic chemical and physical properties in nitrogene opens possibilities in next-generation electronic, optoelectronic, and plasmonic applications.

**Methods**

**Experiments**

Both sample preparation and measurements were conducted in an ultrahigh vacuum system with a base pressure of approximately $1\times10^{-8}$ Pa. The Ag(100) surfaces were prepared through repeated cycles of $Ar^+$ ion sputtering and annealing to ensure cleanliness. Prior to nitrogen deposition, the Ag(100) substrate was maintained at 400 K for 30 min. During the IBAE growth process, low-energy nitrogen ions (~30 eV) were generated using an ion source[41,42] and bombarded onto the clean Ag(100) substrate for 15 min. The pressure of $N_2$ during growth was approximately $3\times10^{-3}$ Pa.

Following the growth process, the samples were characterized *in situ* using LEED, STM, and ARPES. STM imaging was performed in constant-current mode, while STS was carried out with an ac modulation voltage of 30 mV at 667 Hz. All STM images and STS spectra were acquired at 78 K, with bias voltages defined as the sample bias relative to the tip. ARPES measurements were performed using a SPECS PHOIBOS 150 electron energy analyzer and a helium discharge lamp (He Iα, 21.2 eV) at 70 K. XPS measurements were performed at room temperature using an Al Kα X-ray source (SPECS XR50).

**Calculations**

First-principles calculations were conducted based on density functional theory (DFT) using the projector augmented wave method as implemented in the Vienna *Ab initio* Simulation Package (VASP)[43,44]. The generalized gradient approximation (GGA) with the Perdew-Burke-Ernzerhof (PBE) exchange-correlation functional[45,46] was employed. A kinetic energy cutoff of 500 eV was applied to the plane-wave basis set, and a Γ-centered 12×12×1 *k*-point mesh was used for sampling the first Brillouin zone (BZ). The slab model consisted of three layers of the Ag substrate, a monolayer of AgN, a nitrogene layer, and a vacuum layer exceeding 15 Å in thickness. Convergence criteria were set to 0.01 eV/Å for ionic relaxation and $10^{-6}$ eV for the electronic self-consistent loop. The band structure of the slab system was unfolded into the first BZ of the Ag substrate using the orbital-selective band unfolding technique[47].

**Acknowledgements**

We thank Kenya Shimada, Shin-ichiro Ideta, Kumar Yogendra, and Masashi Arita for their support in synchrotron ARPES measurements. This work was supported by the National Key R&D Program of China [Grants No. 2024YFA1408400 (B.F.), 2024YFA1408700 (B.F.), 2021YFA1400502 (K.W.), 2021YFA1202902 (K.W.), and 2024YFA1409100 (L.C.)], the Beijing Natural Science Foundation [Grant No. JQ23001 (B.F.)], the National Natural Science Foundation of China (Grants No. W2411004 (B.F.), 12374197 (B.F.), T2325028 (L.C.), and 12134019 (L.C.)), and the CAS Project for Young Scientists in Basic Research [Grants No. YSBR-047 (B.F.) and YSBR-054 (L.C.)]. We acknowledge the beamtimes in Research Institute for Synchrotron Radiation Science, Hiroshima University (Proposals No. 24AG008 and 24AG009).

## Author contributions

B.F. conceived the research. X.H., Z.G., D.G., D.L., J.G., Q.Z. and Z.C. prepared the samples and performed STM, LEED, and ARPES experiments under the supervision of B.F.; H.C. and H.Z. performed theoretical calculations and analysis under the supervision of S.M.; P.C., L.C., and K.W. contributed to the discussion and interpretation of the experimental data. X.H. and B.F. prepared the manuscript with input from all authors.

## Competing interests

The authors declare no competing interests.

## Data availability

Relevant data supporting the key findings of this study are available within the article and the Supplementary Information file. All raw data generated during the current study are available from the corresponding authors upon request.

**Correspondence and requests for materials** should be addressed to Lan Chen, Kehui Wu, or Baojie Feng.